\documentclass[12pt,preprint]{aastex}

\begin{document}
\title{
Super-solar N/C in the NLS1 Galaxy Markarian\,1044
}
\author{Dale L. Fields\altaffilmark{1}, Smita Mathur\altaffilmark{1}, Richard W. Pogge\altaffilmark{1}, Fabrizio Nicastro\altaffilmark{2} \& Stefanie Komossa\altaffilmark{3}}
\altaffiltext{1}
{Department of Astronomy, the Ohio State University, 140 West 18th
Avenue, Columbus, OH 43210, USA}
\altaffiltext{2}
{SAO,
60 Garden Street, 02138, Cambridge, MA, USA}
\altaffiltext{3}
{Max-Planck-Institut f\"ur extraterrestrische Physik,
Giessenbachstrasse 1, D-85748 Garching, Germany}

\email{fields@astronomy.ohio-state.edu}

\begin{abstract}

Narrow-Line Seyfert 1s (NLS1s) are known have extreme values of a number
of properties compared to Active Galactic Nuclei (AGN) as a class.  In
particular, previous emission-line studies have suggested that NLS1s are
unusually metal rich compared to broad-line AGN of comparable
luminosity.  We present low- and medium-resolution spectroscopic
observations of the NLS1 Markarian\,1044 with the Hubble Space Telescope
Imaging Spectrometer (STIS).  We identify two blueshifted intrinsic
absorption systems at $-1145$ and $-295$\,km\,s$^{-1}$ relative to the
systemic velocity of the galaxy.  Using a simple photoionization model
of the absorbing gas, we find that the strongest and best-measured of
the absorption systems has $N/C\approx6.96\,[N/C]_\sun$.  We also report
on the discovery of three new Ly$\alpha$ forest lines with $\log N_{H
I}\ge12.77$.  This number is consistent with the 2.6 expected in the
path length to the source redshift of Mrk\,1044.

\end{abstract}

\section{Introduction} \label{sec:intro}

The Narrow-Line Seyfert 1 (NLS1) class has come a long way since being
identified by \citet{OP}.  There, they were defined as Seyfert 1s with
relatively narrow permitted lines ($\le2000$\,km\,s$^{-1}$) and a weak
[\ion{O}{3}]/H$\beta$\ ratio ($\le3$).  The principal component analysis
of \citet{BG} found that NLS1s lie at one extreme end of the eigenvector
with the most variation in AGN spectral properties.  In addition, NLS1s
often have very steep X-ray spectra, placing them on one end of the
anti-correlation between soft X-ray spectral slope and H$\beta$\ FWHM
seen in Seyfert 1s \citep{BBF} and quasars \citep{L97}.

While work continues to identify the physical driver of eigenvector 1,
the leading contender is accretion rate expressed as a fraction of the
Eddington accretion rate ($\rm \dot{m}=\dot{M}/\dot{M}_{Edd}$) as
suggested by \citet{P95}.  This would imply that NLS1s and some quasars
are accreting at a large fraction of the Eddington rate.  This receives
support from evidence that NLS1s appear to have smaller supermassive
black holes than many other Seyferts, and yet have comparable
luminosities \citep[and references therein]{G04} and references therein.
Another type of AGN suspected to have a large fractional accretion rate
are the high-redshift quasars.  These have much larger luminosities, and
correspondingly larger black hole masses than other AGN.

One other property that NLS1s and high-redshift quasars appear to share
is super-solar gas-phase metallicity.  The NLS1 PG1404+226 has unusually
strong \ion{N}{5} absorption lines relative to Ly$\alpha$, \ion{C}{4}
and X-ray absorption features, and this can only be explained by
super-solar metallicity \citep{M2Kb}.  \citet{W99} found that the
strengths of the \ion{N}{5}\,$\lambda1240$ emission lines systematically
stronger than average, and the \ion{C}{4}\,$\lambda1549$ were
systematically weaker in AGN with narrow permitted lines.  This is
similar to what appears in high redshift quasars \citep{HF,O94,SH}.  In
addition, the strength of the fluorescent Fe-K$\alpha$ emission line in
some NLS1s is also thought to indicate super-solar abundances
\citep{CJ}.  Finally, \citet{KM01} noted that the steep X-ray spectra
seen in NLS1s eliminates the possibility of having a multi-phase medium
with solar abundances.  Only if the gas is super-solar can it
successfully form a pressure-confined broad-line region.

Metallicity, and more importantly, abundance ratios reflect the star
formation history of the nuclear gas.  This in turn might tell us about
the particular circumstances leading to the activation of the AGN.
There are theories that link the many similarities between NLS1s and
quasars into a scenario stating that these objects are ``young,'' that
is, the black holes have recently begun (re-)accreting gas \citep{M2Ka}.
In the case of Narrow-Line Seyfert 1s, this is very likely, as the mass
e-folding time is short enough to conclude that if that black hole had
any similarly major accretion events in its past, it would not have the
relatively low mass it has today.  There are even suggestions that some
NLS1s do not fall directly on the $M_{BH}-\sigma$ relation
\citep{M01,GM04}.  Because NLS1s could be part of an evolutionary track
(as opposed to the Unified Model), there is no a priori reason why NLS1s
should have the same star formation history, and thus metal abundance
ratios, as normal Seyfert galaxies.  Indeed, \citet{SN} find that NLS1s
have significantly higher \ion{N}{5}/\ion{C}{4} ratios than standard
Seyferts.  This result, however, is not confirmed in
\ion{N}{4}/\ion{C}{4}.  In addition, data on this ratio is based upon
emission line ratios.  A single emission line can originate in multiple
locations, each with different physical conditions, complicating the
conversion of line flux to column densities to abundance.  If one wishes
to get an unambiguous result, one has to measure the absorption lines.
With this in mind, the emission lines results indicate super-solar
metallicity \citep{HF}, but require confirmation.

We seek to determine the metal abundance of a NLS1 using absorption
lines.  The largest contributor to the uncertainty in the abundances is
the uncertainty in the photoionization correction.  This must be well
determined if there is to be any confidence in our results.
Historically, studies have contained few transitions and few (sometimes
only one) species per element due to limited wavelength coverage
(e.g. only UV or only X-ray) and thus do not have the leverage to
determine the correct ionization model.  Our program consists of
near-simultaneous observations of the NLS1 Markarian\,1044 with the HST,
FUSE, and Chandra observatories.  This long wavelength baseline ensures
that we will have the range of species and elements to accurately model
the ionization correction.  In this paper, we present observations of
Mrk\,1044 with the FUV-MAMA instrument onboard the Hubble Space
Telescope.  Future papers will detail the FUSE and Chandra observations
and the self-consistent model derived from all three's observations.  In
\S\ref{sec:data} we detail the the FUV-MAMA observation and data.  We
then follow in \S\ref{sec:analysis} with the analysis.  We discuss the
results in \S\ref{sec:discuss} and give concluding remarks in
\S\ref{sec:conclude}.

\section{Observations and Data Reduction} \label{sec:data}

Mrk\,1044 was observed on UTC 2003 Jun 28 using the Hubble Space
Telescope Imaging Spectrometer (STIS) with the FUV-MAMA detector.  All
observations were acquired during a single 5-orbit visit that ran from
MJD 52818.19912 through MJD 52818.49349.  A journal of observations is
given in Table~\ref{tab:obs}.  The 52X0.2 aperture and a clear filter
were used for all observations, centering the aperture on the bright
nucleus of Mrk\,1044.  Two low-dispersion UV spectra with the G140L
grating (central wavelength of $\lambda1426$\AA) were acquired during
the first orbit, with a cumulative exposure time of 2311\,seconds.
Medium-dispersion spectra were acquired with the G140M grating during
the remaining 4 orbits at central wavelengths of $\lambda1222$,
$\lambda1272$, and $\lambda1567$ covering the emission lines of
Ly$\alpha$, \ion{N}{5}, and \ion{C}{4}, respectively.  The Ly$\alpha$
and \ion{C}{4} regions were observed for 1 orbit each, acquiring 2
spectra per orbit for cumulative exposure times of 2734\,seconds each,
and \ion{N}{5} was observed for 2 orbits (4 spectra) for a cumulative
exposure time of 5583\,seconds.

The standard STSDAS\footnote{STSDAS is a product of the Space Telescope
Science Institute, which is operated by AURA for NASA.} {\tt calstis}
pipeline was used to reduce the raw data to wavelength- and
flux-calibrated 2D spectra.  The nuclear spectra of Mrk\,1044 were
extracted from the linearized 2D spectra (STIS ``x2d'' science frames)
using a 9 pixel (0.45 arcsec) aperture using the {\tt spectroid}
centroiding aperture-extraction task in XVista\footnote{XVista is the
current incarnation of Lick Observatory Vista, maintained in the public
domain at http://ganymede.nmsu.edu/holtz/xvista.}.  Corresponding error
spectra were also extracted from the STIS error arrays for each spectrum
and co-added in quadrature.  These 1D spectra form the basis of our
subsequent analysis.

\begin{figure}
\epsscale{0.9}
\plotone{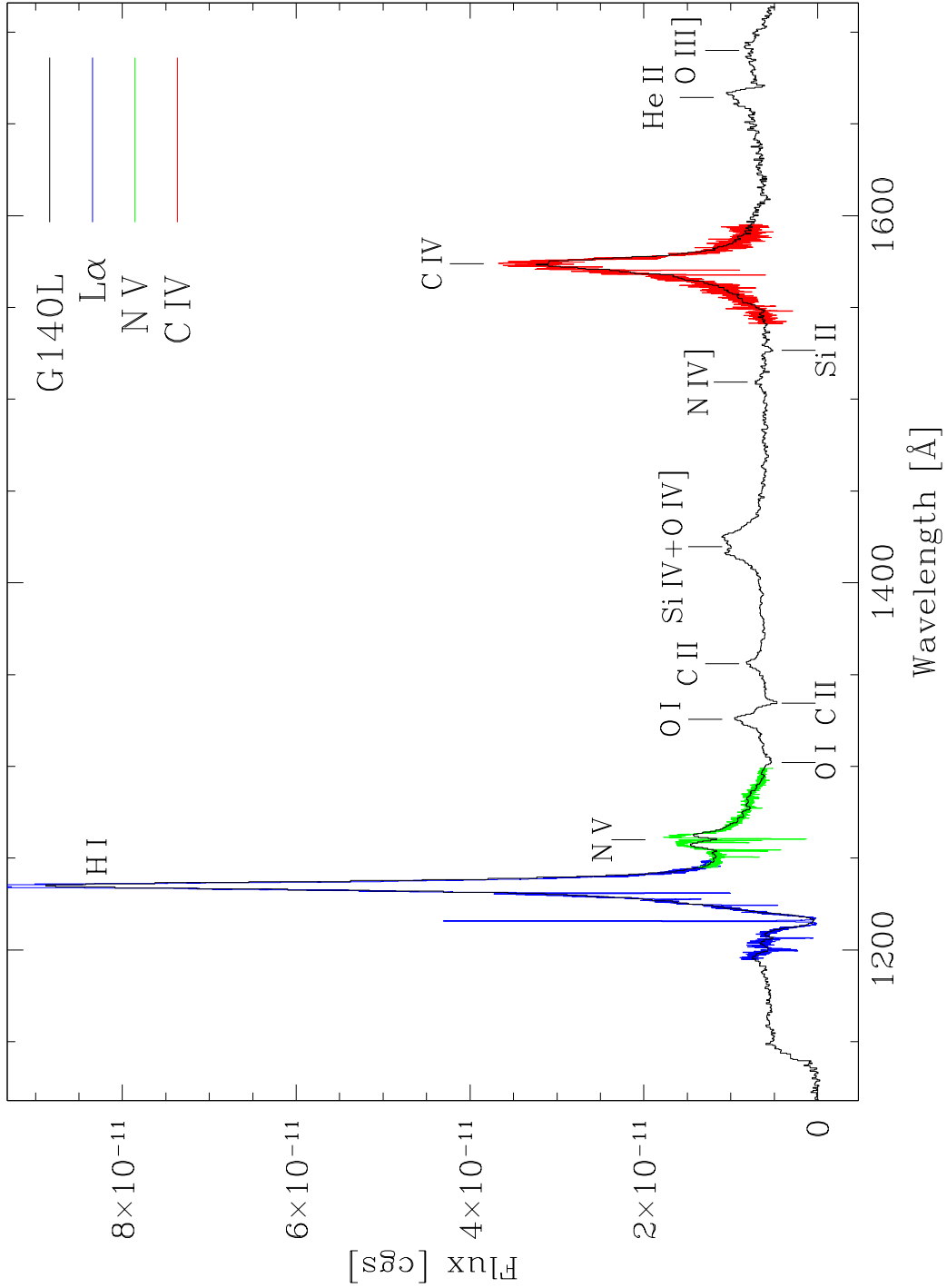}
\caption{\label{fig:masterspec}
The three G140M spectra superposed over the G140L spectrum.  The
necessity of obtaining high-resolution spectra to determine the
metallicity can be seen very well here.  The \ion{C}{4} features are
completely wiped out in the G140L spectrum.  The \ion{N}{5} feature
visible in the G140L spectrum is, after consulting the G140M spectrum,
seen mostly to be caused by Galactic absorption.  The emission lines
originating from Mrk\,1044 are marked above the spectrum.  The
absorption lines caused by Galactic ISM are marked below the
spectrum. }\end{figure}

\begin{figure}
\epsscale{0.95}
\plotone{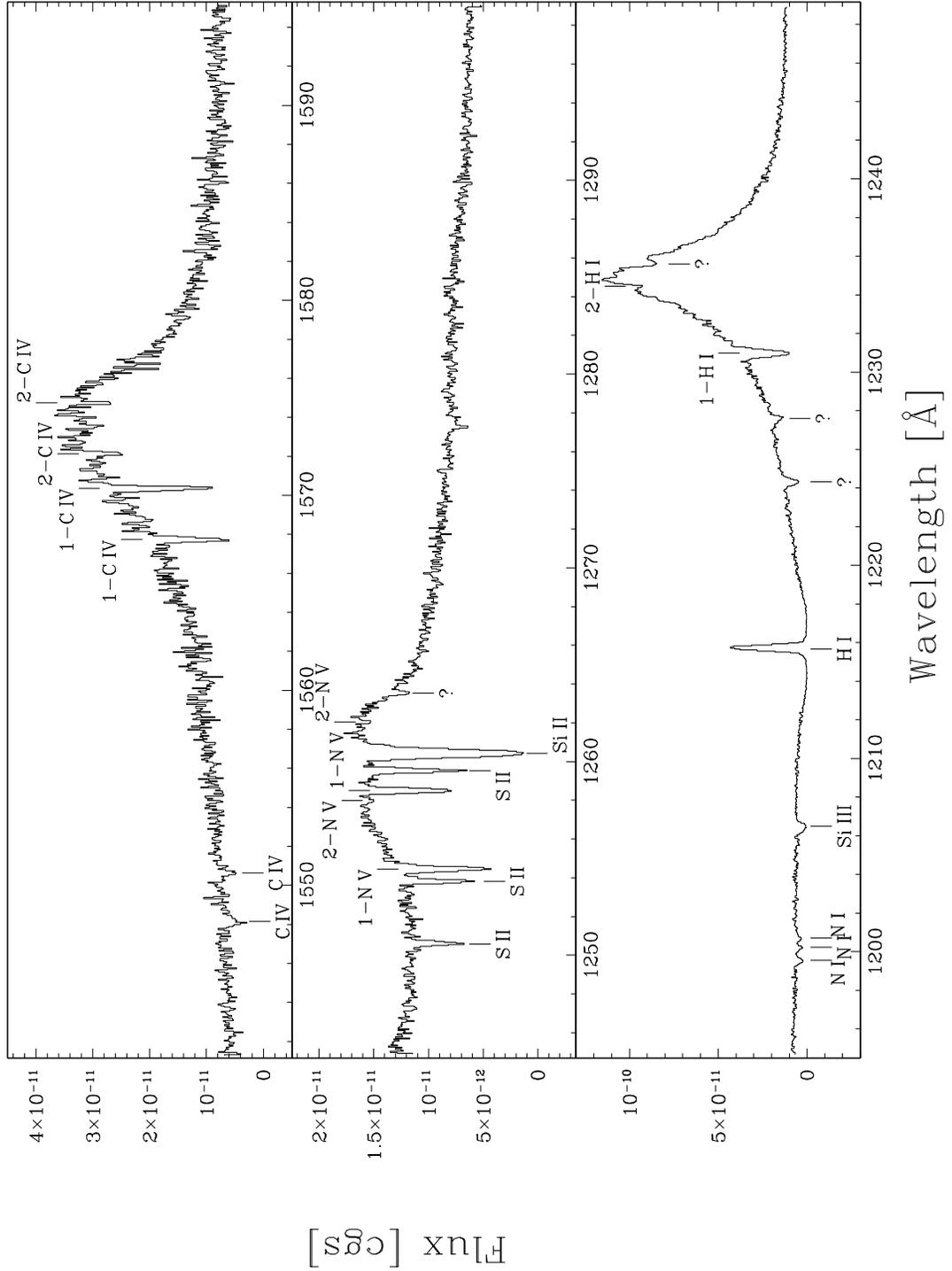}
\caption{\label{fig:threespec}
The three G140M spectra each shown separately.  The absorption systems
intrinsic to Mrk\,1044 are marked above the spectrum and the identified
Galactic lines are marked below the spectrum.  Unidentified lines are
marked as such.}\end{figure}

The G140L spectrum provides us with a reliable internal relative
intensity calibration for the three non-overlapping G140M spectra, and
provides information on the overall UV spectral slope.  The G140L
spectrum with the flux-calibrated G140M spectra superimposed is shown in
Figure~\ref{fig:masterspec}. The individual flux-calibrated G140M
spectra are shown in Figure~\ref{fig:threespec}.

\section{Analysis} \label{sec:analysis}

As can be seen in Figure~\ref{fig:masterspec}, the UV spectrum of
Mrk\,1044 has a flat UV continuum with a slope of $\alpha=0.0$ for
$f_{\nu}\propto\nu^{-\alpha}$.  Its emission line features are similar
to those of many other AGN.  Most prominent are the Ly$\alpha$,
\ion{N}{5}, and \ion{C}{4} emission lines.  Less prominent, but
certainly present are the \ion{Si}{2}\,$\lambda1265$,
\ion{O}{1}\,$\lambda1305$, \ion{C}{2}\,$\lambda1335$,
\ion{Si}{4}\,$\lambda\lambda1394$,$1403$+\ion{O}{4}]\,$\lambda1402$,
\ion{N}{4}]\,$\lambda1486$, \ion{He}{2}\,$\lambda1640$, and
\ion{O}{3}]\,$\lambda1663$ lines.  Emission lines in general have a
narrow ($<1500$\,km\,s$^{-1}$) component and a broad
($>3000$\,km\,s$^{-1}$) base.  The presence of this base causes
significant blending of Ly$\alpha$ and \ion{N}{5}.

The three G140M spectra show a total of 26 possible absorption lines,
all but four of which have good identifications.  Line identification
proceeded from three assumptions: 1) lines near the rest-frame
wavelength of a well-known and strong transition are likely due to
Galactic ISM, 2) any \ion{N}{5} or \ion{C}{4} absorption line systems
intrinsic to Mrk\,1044 will also appear at the same velocity in
Ly$\alpha$, and 3) any remaining unidentified absorption lines blueward
of Mrk\,1044's Ly$\alpha$ emission line, but redward of 1216\AA\ could
be the signature of the low redshift Ly$\alpha$ forest.

\begin{figure}
\epsscale{1.0}
\plotone{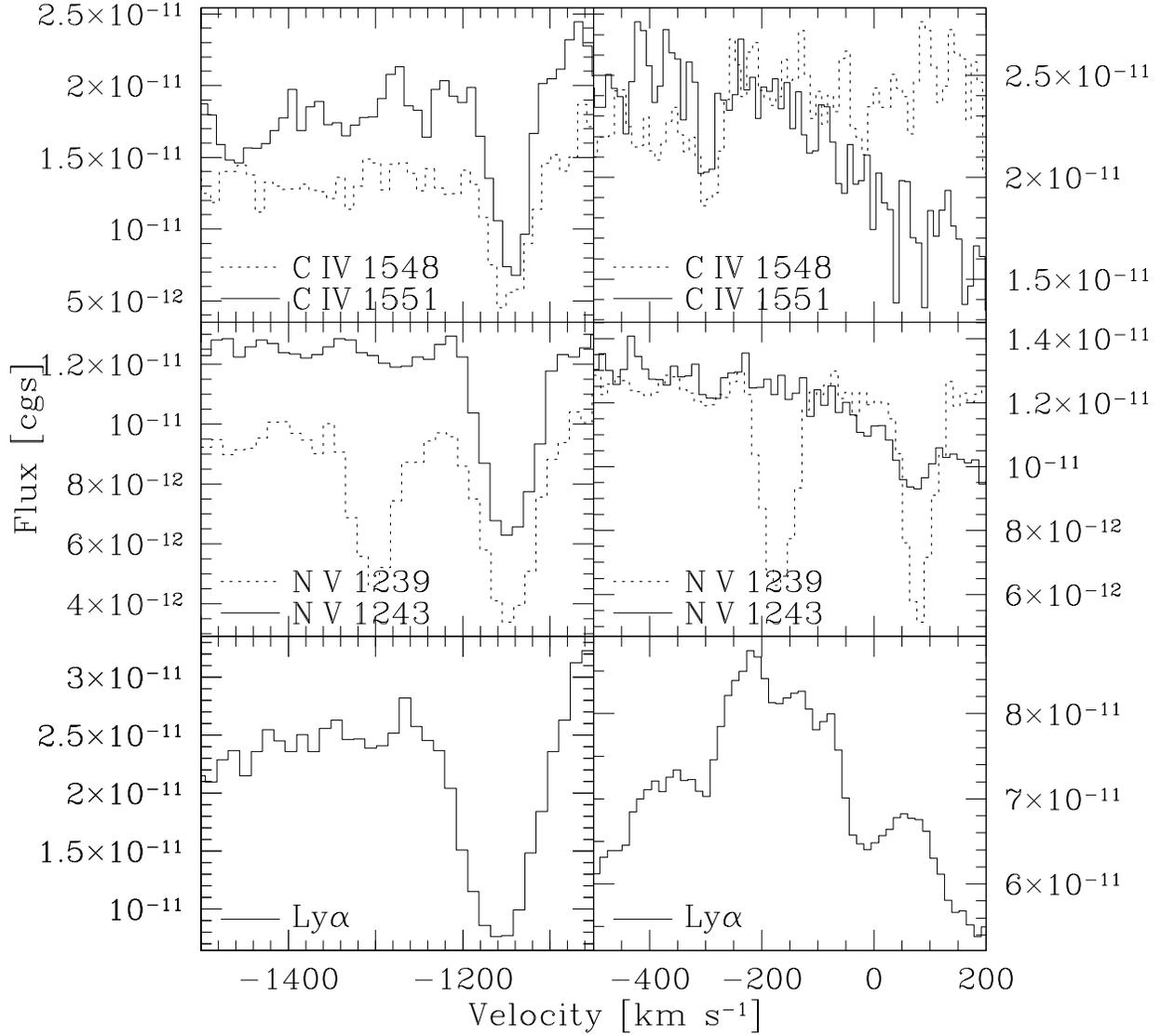}
\caption{\label{fig:velmap}
Velocity maps around Systems 1 (left) and 2 (right).  In the case of
\ion{C}{4} and \ion{N}{5}, the dotted lines signify the shorter
wavelength line of the pair, while the solid lines signify the longer
wavelength line.  System 1 is very secure with all three ions showing
absorption at that velocity.  System 2 is less obvious with the
\ion{N}{5} lines showing little contrast.  Also visible at 
$\sim+75$\,km\,s$^{-1}$ is the chance alignment of a Galactic \ion{S}{2}
line with a yet unidentified line.  Evidence against this candidate
being a true system is the lack of corroborating absorption at that
velocity in \ion{C}{4} and Ly$\alpha$. }\end{figure}

Line doublets of \ion{N}{5} and \ion{C}{4} were identified by
superposing velocity maps for each of the doublets.  From the NASA/IPAC
Extragalactic Database (NED), we adopt a systemic velocity of Mrk\,1044
of 4932\,km\,s$^{-1}$ (z=0.01645).  Velocity maps are shown in
Figure~\ref{fig:velmap} for Ly$\alpha$, \ion{N}{5}, and \ion{C}{4}.
From these, we identify two absorption systems at $-1145$\,km\,s$^{-1}$
and $-295$\,km\,s$^{-1}$.  We find that the two candidate absorption
systems in \ion{N}{5} at $-2100$\,km\,s$^{-1}$ and $+75$\,km\,s$^{-1}$
are actually Galactic absorption from
\ion{S}{2}\,$\lambda\lambda1251$,$1254$ in the case of the former and
\ion{S}{2}\,$\lambda1260$ and an unknown absorption line superposed in
velocity space in the case of the latter.  In both cases, no
corresponding absorption system is detected in \ion{C}{4} or in
Ly$\alpha$.

We use two methods (each with slightly varying analysis paths) to
measure the column densities of the absorbing material.  The first is
the standard curve of growth.  More details of this process may be found
in \citet{Spitzer}.  To determine the velocity spread parameter, we
calculate values of the column density for each line for 'b' values of
10, 20, 40, and 80\,km\,s$^{-1}$.  We then assume the velocity spread
that gives the most consistent result between the line pair.  The second
method is the apparent optical depth method described by \citet{SS91}.
This method is usable in well-resolved systems and thus it will provide
an independent method with which to estimate the column densities.

Lines in each of the spectra were first measured approximately with the
analysis package LINER \citep{LINER}.  Best fit parameters for the
emission lines were then used as initial guesses into the STSDAS
analysis package SPECFIT \citep{SPECFIT}.  To determine the absorption
line parameters via the curve of growth method, LINER was used with a
local continuum parameterized by fitting a fifth order polynomial over a
limited wavelength range (e.g. only on one side of an emission line).  A
polynomial of order five was found to be a good balance between fitting
the intrinsic structure of the continuum and letting noise introduce
unphysical structure into the continuum fitting.  In this manner, the
local continuum and then two or three absorption lines would be fit and
the process repeated for the next absorption feature.  Again, these
results were used as inputs into SPECFIT for final determination of the
equivalent widths.

The analysis of the lines using the apparent optical depth method
proceeded differently.  Because estimating the continuum is the single
largest contributor to the systematic error budget, we performed nine
by-eye spline fits to the continuum around each group of lines.  All
nine were done as reasonable fits, but several were purposefully fit to
the outer envelope of the noise.  A mean was then taken of the nine
resultant column densities with the scatter functioning as our
measurement of error.

\subsection{Mrk\,1044 Emission Lines} \label{sec:mrkemi}

Most of the metallicity studies of AGN have been restricted to using the
emission lines only.  Therefore, we give measurements of our emission
lines to place the results of our absorption line study within the
proper context.  For each ion Table~\ref{tab:emission} gives the rest
wavelength, the observed equivalent width, and the FWHM of the fit
components.  For most emission lines we fit both a narrow and a broad
component.  Both the \ion{N}{5} and \ion{C}{4} emission lines exhibited
an excess of red-ward flux which we parameterized with an additional
Gaussian.  The \ion{Si}{4}+\ion{O}{4}] complex around 1400\AA\ can be
parameterized with several narrow components and one very broad one, but
these are so poorly constrained individually that we only report the
gross properties of the complex.

The literature contains many measures of AGN metallicity.  \citet{SN}'s
Figure 2 gives \ion{N}{5}/\ion{C}{4} and \ion{N}{5}/\ion{He}{2} as a
function of $\nu L_{\nu}$ at 1450\AA.  Mrk\,1044 has ratios of 0.6 and
3.1, respectively, at $1.3\times10^{12}L_\sun$.  These values place
Mrk\,1044 right in the middle of the NLS1 metallicity distributions,
though this object is slightly more luminous than the rest of the NLS1
sample considered by \citet{SN}.  Nevertheless, Mrk\,1044 remains at or
above the mean metallicity-luminosity correlation.  We can also place
Mrk\,1044 alongside the ultraviolet measurements of NLS1s in
\citet{W00}.  This object's \ion{C}{4}/Ly$\alpha$ value of 0.59 is
slightly higher than any of the NLS1s shown in their Figure 2.
Mrk\,1044's (\ion{Si}{4}+\ion{O}{4}])/Ly$\alpha$ value of 0.17 is at the
upper value bounded by their NLS1 sample.  Finally, we can compare this
object against the ``nitrogen-rich'' quasar Q0353-383 \citep{O80} and
its recently discovered fraternity \citep{BO}.  One of the metallicity
indicator used in these studies is $f_{CIV}/(f_{\rm Ly\alpha}+f_{NV})$,
of which Q0353-383 has the lowest value found to date: 0.07.  By
comparison, Mrk\,1044 has a value of 0.45, a factor of six difference.
It should be noted that \citep{O80} found a value of
$N/C\approx10$-$20\,[N/C]_\sun$, which is only a factor of 2-3 higher
than the value we will discuss below in \S\ref{sec:discuss}.

\subsection{Mrk\,1044 Absorption Line Systems} \label{sec:mrkline}

Of the 24 absorption lines found in our UV spectra, 10 of these appear
to be intrinsic to Mrk\,1044, organized into two distinct absorption
systems.  System 1 at $-1145$\,km\,s$^{-1}$ contains strong lines of
Ly$\alpha$, \ion{N}{5}, and \ion{C}{4}.  System 2 at
$-295$\,km\,s$^{-1}$ contains strong lines of Ly$\alpha$ and \ion{C}{4},
but has weak \ion{N}{5}.  Due to their inherent weakness, estimations of
the strengths of the \ion{N}{5} lines in System 2 suffer accordingly.
Table~\ref{tab:mrk1044} gives the measured and derived parameters for
these systems: the observed wavelength, the observed equivalent width,
the column densities derived both the curve of growth method and the
optical depth integration, and the velocity offset from the systemic.
Equivalent widths are typically around 0.2\AA\ in the case of System 1
and 0.02\AA\ in the case of System 2.  The errors in the equivalent
widths were determined by using estimates of the photon noise in each
pixel.  The errors in the velocity are set at 0.5 pixels, as given by
the STIS Instrument Handbook for Cycle 13.  Centroiding errors are much
smaller and therefore ignored.

When we calculate the column densities using the curve of growth method,
we find that a velocity spread parameter of 20\,km\,s$^{-1}$ gives the
most consistent result for both pairs of \ion{N}{5} and \ion{C}{4}
lines.  If we assume that this contributes the greatest to the width of
the line, we can determine if these lines are fully resolved.  The STIS
Instrument Handbook gives a nominal spectral resolution of 1.4 pixels.
This corresponds to between 18 and 14\,km\,s$^{-1}$ in width at the
lines of \ion{N}{5} and \ion{C}{4}, respectively.  This is approximately
similar to the preferred curve of growth width of 20\,km\,s$^{-1}$ so we
will use that as our measure of intrinsic width.  In the case of System
1, the FWHM of the \ion{N}{5} and \ion{C}{4} line pairs are 71, 71, 50
and 47\,km\,s$^{-1}$, from blue to red.  All of these lines are fully
resolved.  In the case of System 2, the FWHM of the lines are 90, 100,
43 and 38\,km\,s$^{-1}$.  The \ion{C}{4} lines are close to being
resolved.  While the best-fit values of the \ion{N}{5} line widths
suggest these two lines are resolved, the large uncertainties prevent
this from being definite.

We calculate the degree of saturation of these two systems by measuring
the ratios of the equivalent widths of each line doublet.  Both
\ion{N}{5} and \ion{C}{4} have natural ratios of 2.0.  This can be used
to derive a covering fraction ($C_f$) for each system following
\citet{H97}.  System 1 has a consistent value of $C_f=0.80$ determined
from both the \ion{N}{5} and \ion{C}{4} lines.  System 2 has less
consistent values with $C_f$ of 0.97 (\ion{N}{5}) and 0.94 (\ion{C}{4}),
but this is not unexpected given the more marginal nature of System 2's
\ion{N}{5} doublet, the $\lambda1243$\AA\ line of which we only detect at
about the 2.5$\sigma$ level.

We calculate \ion{N}{5}/\ion{C}{4} column ratios for Systems 1 and 2.
We use the weighted mean for each species from each doublet.  As
Table~\ref{tab:mrk1044} shows, both methods find very similar results
for the \ion{N}{5} abundance, but the curve of growth method finds a
systematically smaller \ion{C}{4} abundance than the apparent optical
depth method.  This does not affect the \ion{N}{5}/\ion{C}{4} column
ratios at the 3$\sigma$ level.  In System 1, we find
\ion{N}{5}/\ion{C}{4} ratios of $2.7\pm7$\% and $2.0\pm10$\% for the
curve-of-growth and optical depth methods, respectively.  For System 2
we find ratios of $1.0\pm20$\% and $0.7\pm30$\% for the two methods
respectively.

\subsection{Galactic Absorption Lines} \label{sec:galline}

With good certainty, we detect absorption lines of \ion{N}{1},
Ly$\alpha$, \ion{Si}{2}, \ion{Si}{3}, \ion{S}{2}, and \ion{C}{4} in our
medium-resolution spectra and Ly$\alpha$, \ion{O}{1}, \ion{C}{2}, and
\ion{Si}{2} in our low-resolution spectrum from the ISM of our galaxy.
To determine the velocity spread parameter $b$ for the curve of growth
method, we look for the most consistent result between lines of the same
species.  Species for which we have only a single identified line we
report a range of densities corresponding to a range of $b$ of
20-80\,km\,s$^{-1}$.  We also report column densities using the apparent
optical depth method for all lines.  We give the results in
Table~\ref{tab:galactic} in the form of observed wavelength, likely
identification, the candidate's rest wavelength, the velocity width, the
observed equivalent width, the derived column densities for both
methods, and the velocity relative to Mrk\,1044's systematic velocity.
The velocities of the absorption components are scattered about zero
and, in 12 of 14 cases, within three standard deviations of zero.  It
should be noted that there are systematic velocity shifts with respect
to the parent spectrum.  Two of the G140M spectra (Ly$\alpha$ and
\ion{C}{4}) have all their Galactic absorption lines shifted to the
blue, while the other G140M spectrum (\ion{N}{5}) and the G140L spectrum
have their Galactic absorption lines shifted to the red.  This can also
be seen, though not as clearly, in the lines given in
Table~\ref{tab:mrk1044}.  These shifts are not statistically significant
if one accepts the absolute wavelength calibration error of 0.5 pixels.
The errors in the equivalent widths are estimated from the photon noise.

Most of the FWHMs of the lines in the medium-resolution spectra sit
between 50 and 70\,km\,s$^{-1}$ with a few lines
$\gtrsim100\,km\,s^{-1}$ (the Si lines and \ion{C}{4}\,$\lambda$1548).
The uncertainty in the width is not taken from any SPECFIT fit (which
claims a formal precision on the order of one-tenth to one percent error
on any one measurement) but instead from the scatter around the mean
width measurement over many continuum fittings.  This uncertainty is
approximately 5\,km\,s$^{-1}$ for almost all lines.  The \ion{S}{2}
triplet lines have the best agreement amongst themselves, all coming
within one standard deviation of their mean width, 61\,km\,s$^{-1}$.
The \ion{N}{1} lines have a mean width of 70\,km\,s$^{-1}$ and agree to
within one standard deviation.  The Si lines have a mean width of
126\,km\,s$^{-1}$ and agree to within one standard deviation.  The only
real disagreement is from the \ion{C}{4} lines which disagree at greater
than three standard deviations.

The Galactic \ion{C}{4} lines suffer from another problem.  The
velocities (i.e. centroids) of these lines are in great disagreement
with each other.  The uncertainty in the velocity given is that induced
by the absolute wavelength calibration.  Wavelengths relative to one
another in the same spectrum have their uncertainties smaller by a
factor of 2.5 to 5.  This creates a disagreement in the centroids of the
\ion{C}{4} lines of more than $6\sigma$.  In addition, as can be seen
from visual inspection of the spectrum (Figure~\ref{fig:threespec}), the
profiles of these lines are dissimilar.  It is possible that weak
absorption from High Velocity Clouds (HVCs) is preferentially altering
the shape of the stronger \ion{C}{4}\,$\lambda1548$ line.  This could
also explain the large differences in the FWHM of these lines.  This
hypothesis, however, cannot be tested with the current data and must
wait for FUSE confirmations of the existence of HVCs through the
detection (or not) of \ion{O}{6}.

The three absorption lines in the G140L spectrum (\ion{O}{1},
\ion{C}{2}, and \ion{Si}{2}\,$\lambda1527$) have large widths of 920,
560, and 390\,km\,s$^{-1}$, respectively.  All are fully resolved.  It
is possible that what we identify as simply \ion{O}{1} could also be
blended with \ion{S}{1}\,$\lambda1303$.  This would explain the large
width and would eliminate \ion{O}{1}'s large inferred velocity.
Unfortunately, deblending does not return a clean result, so we cannot
make this correction with any certainty and instead elect to omit it.
The large values of inferred widths in the G140L spectrum is somewhat
surprising, given that the low ionization lines seen in the higher
resolution spectra are all so narrow.  The cause of this is not yet
known and we caution anyone from trusting the properties of these lines
at this time.

The four unidentified absorption lines fail at least one of our
criteria.  To be identified, a line must come from a strong transition
of an abundant element without having a sufficiently large velocity
difference.  For example, \ion{Si}{2}\,$\lambda1264.7$\AA\ could be
identified with the line at 1263.5\AA\ but for the fact that this would
give it a relative velocity of $-286$\,km\,s$^{-1}$.  We also use
additional information such as the fact that the line in question has a
width of 52\,km\,s$^{-1}$ which makes it unlikely to come from Si whose
lines average 123\,km\,s$^{-1}$.  It is possible that the absorption
lines near the Ly$\alpha$ emission line of Mrk\,1044 are Ly$\alpha$
whose systems simply do not have the column to appear in
\ion{N}{5} or \ion{C}{4}.

\subsection{Intervening Ly$\alpha$ Absorption?} \label{sec:lyaline}

\begin{figure}
\epsscale{0.8}
\plotone{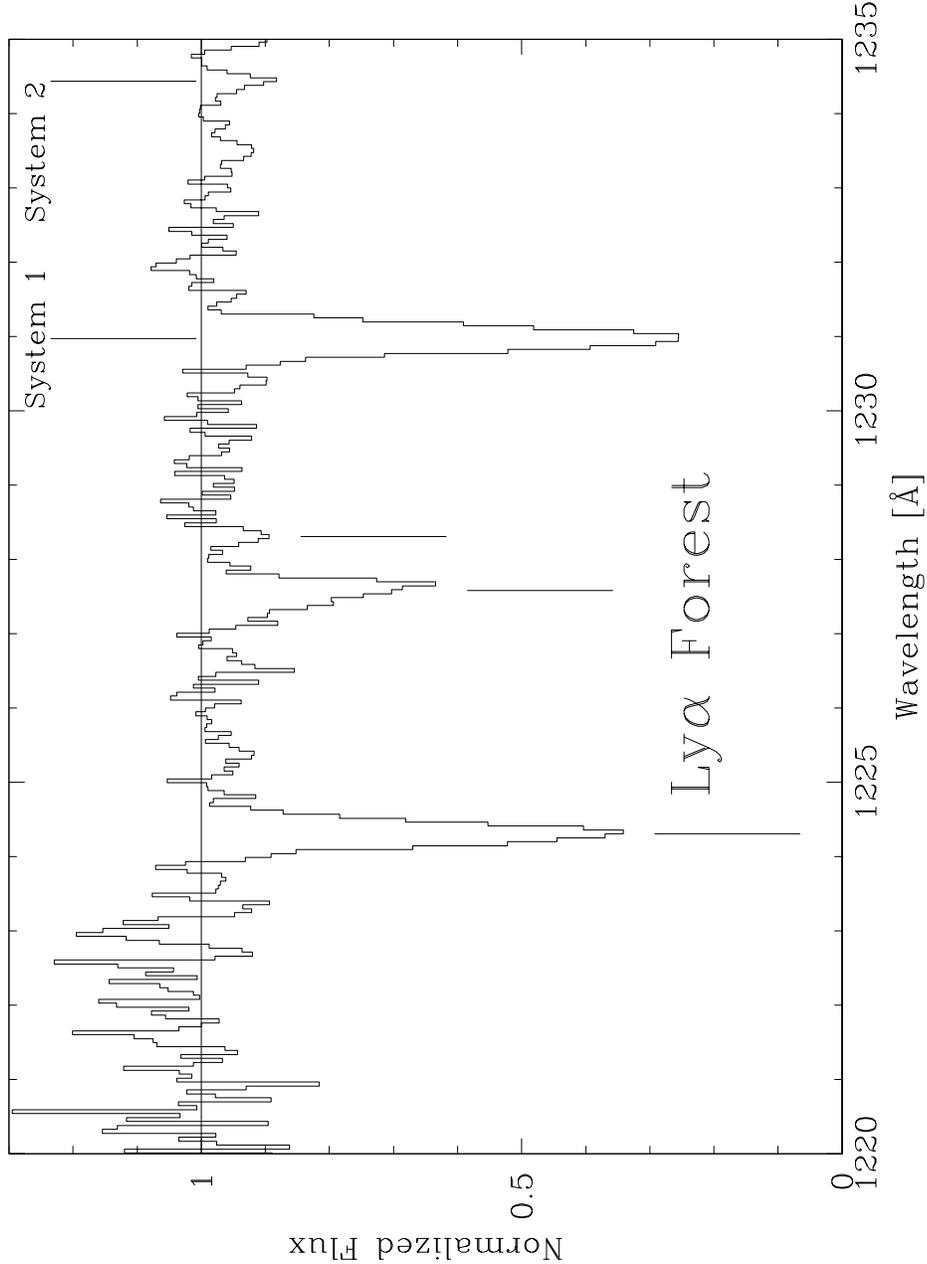}
\caption{\label{fig:lyaforest}
The normalized (continuum divided) spectrum blueward of the Ly$\alpha$
emission line of Mrk\,1044.  This is one of nine by-eye spline fits to
the local continuum (i.e. the Ly$\alpha$ emission line) as described in
\S\ref{sec:analysis}.  The absorption feature at 1228.3\AA\ is a
$3\sigma$ detection.  Other possible absorption features such as those
blueward of 1224\AA\ or at 1233.5\AA\ have significant detections in
only some of the nine continuum fits.  For example, the absorption
feature at 1233.5\AA\ is located near the steepest part of the
Ly$\alpha$ emission line.  Small changes in the perceived ``continuum''
level severely affect its properties.  This also affects the absorption
feature of System 2 to a lesser extent, though System 2 has the
advantage of being confirmed by the presence of \ion{N}{5} and
\ion{C}{4} at the same velocity (see Figure~\ref{fig:velmap}).
}\end{figure}

The two unidentified absorption lines blueward of the Ly$\alpha$
emission line have velocities relative to Mrk\,1044's systemic velocity
of about 2000 and 3000\,km\,s$^{-1}$.  While this does not particularly
favor an origin as material intrinsic to Mrk\,1044, it is not
inconsistent with that hypothesis.  If these absorption lines are due to
intrinsic Ly$\alpha$, the inferred column densities are intermediate
between Systems 1 and 2.  Unlike those two systems, however, these
unidentified systems do not show absorption in either \ion{N}{5} or
\ion{C}{4}.  An alternative hypothesis is that these two systems are due
to the low-redshift Ly$\alpha$-forest.  To test this, we calculate the
number of expected absorption systems from the results of \citet{PSS}.
They parameterize the number of systems per unit log column density and
per unit redshift $\partial^2 N/\partial N_{H I} \partial z$ as $C_{H I}
N_{H I}^{-\beta}$ with $C_{H I}=10^{10.3}$ and $\beta=1.65$ between
$\log N_{H I}$ of 12.3 and 14.5.  We calculate the $3\sigma$ detection
limit of Ly$\alpha$ absorption to be $\log N_{H I}=12.77$.  Because this
is almost a factor of ten lower in column density than the weaker of our
systems, we reanalyze the spectrum for weak lines.  Our best continuum
fit (flattest continuum-divided spectrum) can be seen in
Figure~\ref{fig:lyaforest}.  While any one continuum fit may produce
several well-detected line candidates, we require any line to be
well-detected in the majority of our nine continuum fits.  Only one
additional line passes this test, being detected, on average, at exactly
$3\sigma$.  The wavelengths, velocity widths, equivalent widths,
calculated column densities and relative velocities of the two strong
Ly$\alpha$ systems and the weak, newly found system are given in
Table~\ref{tab:lyaforest}.  For the decade above our detection limit,
and for a path length $\Delta z=z_{\rm Mrk\,1044}=0.01645$, 2.0 systems
are expected, and two systems (within our column density errors) are
detected.  The expected number of systems at all higher column densities
($\log N_{H I}>13.77$) and the same path length is 0.6 and we detect
one.  In this manner, the attribution of these absorption lines to
intergalactic Ly$\alpha$ absorption is perfectly reasonable.

\section{Discussion} \label{sec:discuss}

The \ion{N}{5}/\ion{C}{4} ratio does not tell us the N/C abundance ratio
until we apply an ionization correction for each species.  This
ionization correction is the greatest source of systematic uncertainty
in the determination of abundances.  An accurate ionization correction
requires measurements of multiple species of a single element, and
preferably as many additional elements as possible.  For this reason, we
cannot create an accurate ionization model with just this HST data.  We
are able, however, to set a lower limit on the N/C abundance ratio by
calculating the \ion{N}{5}/\ion{C}{4} ratio in a region with physical
parameters optimized such that the maximum amount of N is in \ion{N}{5}.
We invert the equation

\begin{equation}
\label{eqn:optimal}
\frac{\mbox{\ion{N}{5}}}{\mbox{\ion{C}{4}}}=\frac{A(N)}{A(C)}\times
\frac{\mbox{(\ion{N}{5}/N)}}{\mbox{(\ion{C}{4}/C)}}
\end{equation}

\noindent where A(N)/A(C) is the total Nitrogen to total Carbon
abundance ratio (henceforth N/C), the quantity we want to measure.
Following \citet{HF} (their Figure 10), we find that \ion{N}{5} has a
maximum ionization fraction of 0.40 at $\log U$ of $-1.55$.  The
fraction of carbon in \ion{C}{4} at the same ionization parameter is
0.31.  Using our observed ratio of \ion{N}{5}/\ion{C}{4}=2.67 from
System 1, this gives a N/C ratio of 2.07, or 6.96\,$[N/C]_\sun$
($\pm7$\%).  Doing the same analysis with the weaker absorption lines of
System 2 gives a super-solar N/C ratio of only 2.3\,$[N/C]_\sun$
($\pm20$\%).  The errors quoted only incorporate uncertainties due to
measurements derived from the spectra.

To demonstrate the discrepancies caused by improper ionization
corrections, we compute N/C for two alternative models.  We do this by
running the photoionization-equilibrium code Cloudy\,94\footnote{Cloudy
version C94.00, obtained from the Cloudy webpage
http://www.nublado.org/} \citep{cloudy} with two differing spectral
energy distributions (SEDs): a power law SED with $\alpha=-1.5$ and
Cloudy's standard AGN SED.  We find that System 1 has an abundance ratio
of 4.3\,$[N/C]_\sun$ in the case of the simple power law SED and
2.9\,$[N/C]_\sun$ in the case of the standard AGN template SED.  This
can be compared with value of 6.96 using \citet{HF} as above.  The
estimated error due to photon noise is only $\sim7$\%, which
demonstrates that our result is dominated by systematic uncertainty in
the ionization corrections, and more accurate results await data at
other wavebands that should permit calculation of a detailed
photoionization model for these spectra.  Until such a model is
constructed, we are limited to stating only that Mrk\,1044 has line
strengths consistent with a super-solar N/C ratio.

\section{Conclusions} \label{sec:conclude}

We have analyzed FUV-MAMA spectra of the Narrow-Line Seyfert 1 galaxy
Mrk\,1044.  We find two absorption systems blueshifted with respect to
the systemic velocity of the galaxy.  We find absorption lines from
Galactic ISM in both the low-resolution and medium-resolution spectra.
We find super-solar N/C in the outflow systems of Mrk\,1044 utilizing a
very simple ionization model.  We defer constructing a more detailed
photoionization model until after analyzing the FUSE and Chandra data so
as to avoid the problems that have historically plagued models derived
from single wavelength regions.

\begin{acknowledgements}
The authors wish to thank Mike Crenshaw for his assistance with the
apparent optical depth method.  This research has made use of the
NASA/IPAC Extragalactic Database (NED) which is operated by the Jet
Propulsion Laboratory, California Institute of Technology, under
contract with the National Aeronautics and Space Administration.
Primary support for this work was provided by NASA grant HST-GO-09687.O2
from the Space Telescope Science Institute, which is operated by the
Association of Universities for Research in Astronomy, Inc., under NASA
contract NAS 5-26555.
\end{acknowledgements}

\clearpage

\begin{deluxetable}{ccccc}
\tablecaption{Journal of Observations\label{tab:obs}}
\tablehead{
\colhead{}
&\colhead{Central}
&\colhead{UTC Date}
&\colhead{Exposure}
&\colhead{}\\
\colhead{Grating}
&\colhead{Wavelength [\AA]}
&\colhead{(start)}
&\colhead{Times [s]}
&\colhead{Datasets}}
\startdata
G140L&1425&2003-06-28T04:46:49&1200 \& 1011&O8K401010-20 \\
G140M&1567&2003-06-28T06:14:08&1294 \& 1440&O8K040130-40 \\
     &1222&2003-06-28T07:50:11&1294 \& 1440&O8K040150-60 \\
     &1272&2003-06-28T09:26:15&1294, 1409 \& 2$\times$1440&O8K040170-a0 \\
\enddata
\end{deluxetable}

\begin{deluxetable}{ccccc}
\tablecaption{Emission Lines
\label{tab:emission}}
\tablehead{
\colhead{Ion}
&\colhead{$\lambda_{Rest}$}
&\colhead{Equivalent}
&\colhead{FWHM-Narrow}
&\colhead{FWHM-Broad}\\
\colhead{}
&\colhead{[\AA]}
&\colhead{Width [\AA]}
&\colhead{[\,km\,s$^{-1}$]}
&\colhead{[\,km\,s$^{-1}$]}}
\startdata
\ion{H}{1}&$1216$&$113.6\pm0.1$&$1120.6\pm0.3$&$3406\pm3$ \\
\ion{N}{5}&$1239$,$1243$&$40.2\pm0.4$&$1070\pm40$&$5530\pm50$\,\tablenotemark{a} \\
\ion{O}{1}&$1302$&$4.9\pm0.7$&$1350\pm7$&$3600\pm200$ \\
\ion{C}{2}&$1336$&$2.94\pm0.04$&$1020\pm20$&$3110\pm50$ \\
\ion{Si}{4}+\ion{O}{4}]&$1394$,$1403$+$1402$&$19.19\pm0.03$&$\sim2000$&$\sim10000$ \\
\ion{N}{4}]&$1486$&$0.984\pm0.005$&$1262\pm4$&\nodata \\
\ion{C}{4}&$1548$,$1551$&$67.0\pm0.3$&$1317\pm3$&$3400\pm30$\,\tablenotemark{b} \\
\ion{He}{2}&$1640$&$12.9\pm0.5$&$1440\pm50$&$7500\pm200$ \\
\ion{O}{3}]&$1663$&$5.2\pm0.4$&\nodata&$3400\pm200$ \\
\enddata
\tablenotetext{a}
{\mbox{Line asymmetry requires a third (red) Gaussian with a width of $3450\pm20$}}
\tablenotetext{b}
{\mbox{Line asymmetry requires a third (red) Gaussian with a width of $7320\pm30$}}
\end{deluxetable}

\begin{deluxetable}{ccccccc}
\tablecaption{Measured and Calculated Parameters of Mrk\,1044
\label{tab:mrk1044}}
\tablehead{
\colhead{Line}
&\colhead{Observed}
&\colhead{FWHM}
&\colhead{Equivalent}
&\colhead{log(Column)}
&\colhead{log(Column)}
&\colhead{Velocity}\\
\colhead{}
&\colhead{Wavelength [\AA]}
&\colhead{[\,km\,s$^{-1}$]}
&\colhead{Width [m\AA]}
&\colhead{[$cm^{-2}$]\,\tablenotemark{a}} 
&\colhead{[$cm^{-2}$]\,\tablenotemark{b}}
&\colhead{[\,km\,s$^{-1}$]}}
\startdata
System 1&&&&& \\
Ly$\alpha$&$1230.9819$&$92\pm7$&$341\pm5$&$15.07^{+0.05}_{-0.05}$&$14.11^{+0.04}_{-0.04}$&$-1156\pm7$ \\
\ion{N}{5}$1239$&$1254.4633$&$71\pm3$&$209\pm8$&$14.38^{+0.05}_{-0.04}$&$14.33^{+0.03}_{-0.03}$&$-1143\pm6$ \\
\ion{N}{5}$1243$&$1258.4999$&$71\pm3$&$162\pm6$&$14.44^{+0.03}_{-0.03}$&$14.45^{+0.02}_{-0.02}$&$-1143\pm6$ \\
\ion{C}{4}$1549$&$1567.7329$&$50\pm6$&$204\pm5$&$13.94^{+0.02}_{-0.03}$&$14.01^{+0.06}_{-0.07}$&$-1147\pm5$ \\
\ion{C}{4}$1551$&$1570.3602$&$47\pm5$&$156\pm4$&$14.04^{+0.02}_{-0.02}$&$14.33^{+0.05}_{-0.06}$&$-1143\pm5$ \\
System 2&&&&& \\
Ly$\alpha$&$1234.4395$&$50\pm8$&$16\pm3$&$12.48^{+0.07}_{-0.08}$&$12.63^{+0.08}_{-0.10}$&$-303\pm6$ \\
\ion{N}{5}$1239$&$1257.9923$&$90\pm20$&$23\pm5$&$13.07^{+0.09}_{-0.11}$&$13.18^{+0.11}_{-0.14}$&$-290\pm6$ \\
\ion{N}{5}$1243$&$1262.0425$&$100\pm50$&$10\pm4$&$13.00^{+0.15}_{-0.21}$&$12.99^{+0.25}_{-0.63}$&$-289\pm6$ \\
\ion{C}{4}$1548$&$1572.1322$&$43\pm10$&$42\pm4$&$13.03^{+0.04}_{-0.04}$&$13.23^{+0.10}_{-0.13}$&$-295\pm5$ \\
\ion{C}{4}$1551$&$1574.7397$&$38\pm10$&$29\pm2$&$13.16^{+0.03}_{-0.03}$&$13.44^{+0.11}_{-0.14}$&$-296\pm5$ \\
\enddata
\tablenotetext{a}
{Derived from curve of growth arguments}
\tablenotetext{b}
{Derived from optical depth integration}
\end{deluxetable}

\begin{deluxetable}{cccccccc}
\tablecaption{Galactic and Unidentified Absorption Lines
\label{tab:galactic}}
\tablehead{
\colhead{$\lambda_{Obs}$}
&\colhead{Ion}
&\colhead{$\lambda_{Rest}$}
&\colhead{FWHM}
&\colhead{Equivalent}
&\colhead{log(Column)}
&\colhead{log(Column)}
&\colhead{Velocity}\\
\colhead{[\AA]}
&\colhead{}
&\colhead{[\AA]}
&\colhead{[\,km\,s$^{-1}$]}
&\colhead{Width [m\AA]}
&\colhead{[$cm^{-2}$]\,\tablenotemark{a}}
&\colhead{[$cm^{-2}$]\,\tablenotemark{b}}
&\colhead{[\,km\,s$^{-1}$]}}
\startdata
$1199.5251$&\ion{N}{1}&$1199.550$&$76\pm7$&$210\pm10$&$14.50^{+0.46}_{-0.34}$&$14.25^{+0.06}_{-0.07}$&$-6\pm7$ \\
$1200.1924$&\ion{N}{1}&$1200.223$&$65\pm5$&$160\pm10$&$14.41^{+0.36}_{-0.35}$&$14.30^{+0.05}_{-0.06}$&$-8\pm7$ \\
$1200.6842$&\ion{N}{1}&$1200.710$&$90\pm20$&$170\pm10$&$14.76^{+0.40}_{-0.36}$&$14.50^{+0.05}_{-0.06}$&$-6\pm7$ \\
$1206.4296$&\ion{Si}{3}&$1206.500$&$140\pm10$&$500\pm20$&13.53-16.72&$13.63^{+0.03}_{-0.03}$&$-18\pm7$ \\
$1224.3067$\,\tablenotemark{c}&\nodata&$$&$88\pm4$&$251\pm9$&$$&$$&$$ \\
$1227.6204$\,\tablenotemark{c}&\nodata&$$&$122\pm10$&$175\pm7$&$$&$$&$$ \\
$1235.6051$&\nodata&$$&$82\pm9$&$42\pm3$&$$&$$&$$ \\
$1250.6094$&\ion{S}{2}&$1250.584$&$60\pm6$&$93\pm8$&$15.15^{+0.04}_{-0.04}$&$15.18^{+0.05}_{-0.05}$&$+6\pm6$ \\
$1253.8245$&\ion{S}{2}&$1253.811$&$60\pm5$&$129\pm7$&$15.05^{+0.03}_{-0.03}$&$15.09^{+0.03}_{-0.04}$&$+3\pm6$ \\
$1259.5310$&\ion{S}{2}&$1259.519$&$64\pm5$&$169\pm5$&$15.10^{+0.02}_{-0.02}$&$15.13^{+0.02}_{-0.03}$&$+3\pm6$ \\
$1260.4310$&\ion{Si}{2}&$1260.422$&$125\pm3$&$498\pm8$&13.74-16.62&$13.24^{+0.02}_{-0.02}$&$+2\pm6$ \\
$1263.5315$&\nodata&$$&$49\pm9$&$33\pm6$&$$&$$&$$ \\
$1303.0942$\,\tablenotemark{d}&\ion{O}{1}&$1302.168$&$883\pm7$&$553\pm5$&15.04-18.45&$14.80^{+0.08}_{-0.09}$&$+210\pm70$ \\
$1334.9145$\,\tablenotemark{d}&\ion{C}{2}&$1334.532$&$560\pm10$&$720\pm30$&14.82-20.81&$14.18^{+0.21}_{-0.42}$&$+90\pm70$ \\
$1527.0709$\,\tablenotemark{d}&\ion{Si}{2}&$1526.707$&$380\pm60$&$280\pm50$&14.09-14.50&$14.09^{+0.08}_{-0.10}$&$+70\pm60$ \\
$1548.1517$&\ion{C}{4}&$1548.187$&$97\pm8$&$240\pm20$&$13.83^{+0.04}_{-0.05}$&$13.79^{+0.09}_{-0.11}$&$-7\pm5$ \\
$1550.6375$&\ion{C}{4}&$1550.772$&$50\pm10$&$120\pm20$&$13.80^{+0.05}_{-0.06}$&$13.76^{+0.08}_{-0.10}$&$-26\pm5$ \\
\enddata
\tablenotetext{a}
{Derived from curve of growth arguments}
\tablenotetext{b}
{Derived from optical depth integration}
\tablenotetext{c}
{Possible Ly$\alpha$ forest lines.  See Table~\ref{tab:lyaforest}.}
\tablenotetext{d}
{From the low-resolution spectrum.  The unusually large FWHMs indicate that these values should be used with caution.}
\end{deluxetable}

\begin{deluxetable}{ccccccc}
\tablecaption{Likely Ly$\alpha$ Forest Lines
\label{tab:lyaforest}}
\tablehead{
\colhead{$\lambda_{Obs}$}
&\colhead{FWHM}
&\colhead{Equivalent}
&\colhead{log(Column)}
&\colhead{log(Column)}
&\colhead{Rest Frame}
&\colhead{Mrk\,1044}\\
\colhead{[\AA]}
&\colhead{[\,km\,s$^{-1}$]}
&\colhead{Width [m\AA]}
&\colhead{[$cm^{-2}$]\,\tablenotemark{a}}
&\colhead{[$cm^{-2}$]\,\tablenotemark{b}}
&\colhead{Velocity [\,km\,s$^{-1}$]}
&\colhead{Velocity [\,km\,s$^{-1}$]}}
\startdata
$1224.3067$&$88\pm4$&$251\pm9$&$14.27^{+0.07}_{-0.07}$&$13.80^{+0.02}_{-0.03}$&$2130\pm7$&$-2801\pm7$ \\
$1227.6204$&$120\pm10$&$175\pm7$&$13.79^{+0.04}_{-0.04}$&$13.56^{+0.05}_{-0.06}$&$2937\pm7$&$-1994\pm7$ \\
$1228.3089$&$70\pm20$&$45\pm15$&$12.96^{+0.14}_{-0.19}$&$12.77^{+0.10}_{-0.13}$&$3117\pm7$&$-1814\pm7$ \\
\enddata
\tablenotetext{a}
{Derived from curve of growth arguments}
\tablenotetext{b}
{Derived from optical depth integration}
\end{deluxetable}

\begin{thebibliography}{}

\bibitem[Bentz \& Osmer(2004)]{BO}
Bentz, M. \& Osmer, P.  2004, \aj, 127, 576

\bibitem[Boller, Brandt \& Fink(1996)]{BBF}
Boller, Th., Brandt, N., \& Fink, H.  1996, A\&A, 305, 53

\bibitem[Boroson \& Green(1992)]{BG}
Boroson, T. \& Green, R.  1992, \apjs, 80, 109

\bibitem[Brandt \& Boller(1998)]{BB}
Brandt, N. \& Boller, Th.  1998, A\&A, 319, 7

\bibitem[Collin \& Joly(2000)]{CJ}
Collin, S. \& Joly, M.  2000 \nar, 44, 531

\bibitem[Ferland et al.(1998)]{cloudy}
Ferland, G. J., Korista, K. T., Verner, D. A., Ferguson, J. W., Kingdon, J. B., Verner, E. M.  1998, PASP, 110, 761.

\bibitem[Grupe \& Mathur(2004)]{GM04}
Grupe, D. \& Mathur, S. 2004, \apj, 606L, 41

\bibitem[Grupe et al.(2004)]{G04}
Grupe, D., Wills, B.J., Leighly, K.M., \& Meusinger, H.  2004 \aj, 127, 156

\bibitem[Hamann et al.(1997)]{H97}
Hamann, F., Barlow, T.A., Junkkarinen, V., Burbidge, E.M.  1997, \apj, 478, 80

\bibitem[Hamann \& Ferland(1999)]{HF}
Hamann, F. \& Ferland, G.  1991, \araa, 37, 487

\bibitem[Komossa \& Mathur(2001)]{KM01}
Komossa, S. \& Mathur, S.  2001, A\&A, 974, 914

\bibitem[Kriss(1994)]{SPECFIT}
Kriss, G.A.  1994 in Astronomical Data Analysis Software \& Systems III, A.S.P. Conf. Series, Vol. 61, ed. D. R. Crabtree, R. J. Hanisch, \& J. Barnes (Astronomical Society of the Pacific: San Francisco), p. 437.

\bibitem[Laor et al.(1997)]{L97}
Laor, A., Fiore, F., Elvis, M., Wilkes, B., McDowell, J.  1997, \apj, 477, 93

\bibitem[Mathur(2000a)]{M2Ka}
Mathur, S.  2000, \mnras, 314, L17

\bibitem[Mathur(2000b)]{M2Kb}
Mathur, S.  2000, New Astronomy Reviews, 44, 469

\bibitem[Mathur et al.(2001)]{M01}
Mathur, S., Matt, G., Green, P.J., Elvis, M., Singh, K.P.  2001, \apj, 551L, 13

\bibitem[Osmer et al.(1994)]{O94}
Osmer, P., Porter, A., Green, R.  1994, \apj, 436, 678

\bibitem[Osmer(1980)]{O80}
Osmer, P.  1980, \apj, 237, 666

\bibitem[Osterbrock \& Pogge(1985)]{OP}
Osterbrock, D.E. \& Pogge, R.W.  1985, \apj, 297, 166

\bibitem[Penton et al.(2004)]{PSS}
Penton, S.V., Stocke, J.T., \& Shull, J.M.  2004,  \apjs, 152, 29

\bibitem[Pogge \& Owen(1993)]{LINER}
Pogge, R.W., \& Owen, J.M. 1993, LINER; An Interactive Spectral Line Analysis Program, OSU Internal Report 93-01

\bibitem[Pounds, Done \& Osborne(1995)]{P95}
Pounds, K., Done, C., \& Osborne, J.  1995, \mnras, 277, L5


\bibitem[Savage \& Sembach(1991)]{SS91}
Savage, B.D., \& Sembach, K.R.  1991, \apj, 379, 245

\bibitem[Shemmer \& Netzer(2002)]{SN}
Shemmer, O. \& Netzer, H.  2002, \apj, 567, L22

\bibitem[Shields \& Hamann(1997)]{SH}
Shields, J. \& Hamann, F.  1997, RMxAC, 6, 221

\bibitem[Spitzer(1978)]{Spitzer}
Spitzer, L.  1987, Physical Processes in the Interstellar Medium (New
York: Wiley-Interscience), 46


\bibitem[Wills et al.(1999)]{W99}
Wills, B. et al.  1999 in ``Quasars and Cosmology'', Ed. G. Ferland \&
J. Baldwin

\bibitem[Wills et al.(2000)]{W00}
Wills, B.J., Shang, Z. \& Yuan, J.M.  2000, \nar, 44, 511

\end{thebibliography}
\end{document}